\documentclass{aps2010} 
\usepackage{ifpdf}
\ifpdf
  \usepackage[pdftex]{hyperref}
\else
  \usepackage[ps2pdf,colorlinks=true]{hyperref}
\fi

\usepackage{graphicx}
\usepackage{subfigure}
\usepackage{bm}
\usepackage{amsmath}
\usepackage[numbers,square,comma]{natbib}

\title{Intrinsic advantages of the $\bm{w}$ component and spherical imaging\\
  for wide-field radio interferometry}

\author[org1]{\textbf{\emph{\underline{Jason D. McEwen}}}}
\author[org2]{\textbf{\emph{Yves Wiaux}}}

\address[org1]{Institute of Electrical Engineering, 
    Ecole Polytechnique F{\'e}d{\'e}rale de Lausanne (EPFL),\\ Lausanne 1015,
    Switzerland\\
  jason.mcewen@epfl.ch}
\address[org2]{Institute of Electrical Engineering, 
  Ecole Polytechnique F{\'e}d{\'e}rale de Lausanne (EPFL),\\ Lausanne 1015, Switzerland\\
  Institute of Bioengineering, 
  Ecole Polytechnique F{\'e}d{\'e}rale de Lausanne (EPFL),\\ Lausanne 1015, Switzerland\\
  Department of Radiology and Medical Informatics, University of Geneva (UniGE),\\ Geneva 1211,
  Switzerland\\
  yves.wiaux@epfl.ch}

\newcommand{\eqn}[1]{(#1)}

\newcommand{\fig}[1]{Fig.~#1}

\newcommand{\ie}{\mbox{\it i.e.}}

\newcommand{\ska}{{SKA}}
\newcommand{\skatext}{Square Kilometre Array}

\newcommand{\spcend}{\ensuremath{\:}}

\newcommand{\vect}[1]{\ensuremath{\mbox{\boldmath ${#1}$}}}

\newcommand{\el}{\ensuremath{\ell}}

\newcommand{\sind}{\ensuremath{{\rm s}}}

\newcommand{\vis}{\ensuremath{y}}

\newcommand{\im}{\ensuremath{x}}

\newcommand{\nim}{\ensuremath{n}}
\newcommand{\visvect}{\ensuremath{\vect{\vis}}}
\newcommand{\imvect}{\ensuremath{\vect{\im}}}

\newcommand{\impvect}{\ensuremath{\vect{\imp}}}
\newcommand{\imsvect}{\ensuremath{\vect{\ims}}}

\newcommand{\nimvect}{\ensuremath{\vect{\nim}}}

\newcommand{\bw}{\ensuremath{w}}

\newcommand{\sparmat}{\ensuremath{\Psi}}
\newcommand{\sensmat}{\ensuremath{\Phi}}

\newcommand{\sensmatp}{\ensuremath{\Phi_{\pind}}}
\newcommand{\sensmats}{\ensuremath{\Phi_{\sind}}}

\newcommand{\opproj}{\ensuremath{\mathbfss{P}}}

\renewcommand{\opproj}{\ensuremath{\mathbf{P}}}
\renewcommand{\sensmatp}{\ensuremath{\Phi}}
\renewcommand{\impvect}{\ensuremath{\vect{\im}}}
\renewcommand{\imsvect}{\ensuremath{\vect{\im}_{\sind}}}

\begin{document}%
\maketitleblock  
\begin{abstract}
  Incorporating wide-field considerations in interferometric imaging
  is of increasing importance for next-generation radio
  telescopes. Compressed sensing techniques for
  interferometric imaging have been extended to wide fields recently,
  recovering images in the spherical coordinate space in which they
  naturally live. We review these techniques, highlighting: (i) how
  the effectiveness of the spread spectrum phenomenon, due to the \bw\
  component inducing an increase of measurement incoherence, is
  enhanced when going to wide fields; and (ii) how sparsity is
  reduced by recovering images directly on the sphere. Both of these
  properties act to improve the quality of reconstructed
  images.
\end{abstract}

\section{Introduction}

We are entering a new era of radio astronomy, with a new generation of
radio interferometric telescopes under construction and
design. Next-generation radio interferometers, such as the
\skatext\footnote{\url{http://www.skatelescope.org/}} (\ska)
\citep{carilli:2004}, will inherently observe very large fields of
view.  Wide fields introduce two important distinctions to standard
interferometric imaging: firstly, interferometric images are
inherently spherical and planar projections necessarily introduce
distortions; and, secondly, non-zero baseline components in the
pointing direction of the telescope, referred to
as \bw\ components, must be taken into account,
introducing a non-negligible modulation of the underlying image.  If
these contributions are ignored, the images recovered from forthcoming
telescopes will not reach their full potential.

The theory of compressed sensing has been applied recently to recover
images from simulated measurements taken by radio interferometric
telescopes \cite{wiaux:2009:cs,wiaux:2009:ss}.  In these works the
effectiveness and flexibility of the compressed sensing approach are
demonstrated, resulting in reconstructed interferometric images of
superior quality to standard methods.  The first work is focused on
compressed sensing for reconstruction \cite{wiaux:2009:cs}, while the
second is focused at the level of acquisition on the spread spectrum
phenomenon, due to the \bw\ component \cite{wiaux:2009:ss} (these
techniques are reviewed in a separate article of the current
proceedings \cite{scaife:2011}).  Furthermore, compressed sensing
techniques have been developed to successfully extract astronomical
signals of interest from interferometric observations corrupted by
background contributions \cite{wiaux:2010:csstring}.  More recently,
compressed sensing interferometric imaging techniques have been
generalised to a wide field-of-view \cite{mcewen:2010:riwfov}.  In
this setting images are recovered directly on the sphere, rather than
a tangent plane.  Contrary to standard interferometric imaging, this approach provides intrinsic advantages that may be exploited in a
compressed sensing framework.

\section{Spherical radio interferometric imaging}

Standard interferometric imaging involves recovering an image from
noisy and incomplete Fourier measurements.  The resulting ill-posed
inverse problem is described by the linear system
\begin{equation}
\label{eqn:vis_linear_plane}
\visvect = \sensmatp \impvect + \nimvect
\spcend ,
\end{equation}
where the linear measurement operator \sensmatp\ relates the
underlying image \impvect\ to the incomplete Fourier measurements
taken by the interferometer \visvect, in the presence of noise
\nimvect.
The measurement operator incorporates the primary beam of the
telescope, the \bw\ component modulation responsible for the spread
spectrum phenomenon \cite{wiaux:2009:ss}, the Fourier transform and a
masking which encodes the incomplete measurements taken by the
interferometer.
In the context of compressed sensing, this problem has been solved by
applying a prior on the sparsity of the signal in a sparsifying basis
\sparmat\ or in the magnitude of its gradient.  The underlying image
is recovered by solving the Basis Pursuit denoising problem
\begin{equation}
\label{eqn:min_bp}
\vect{\alpha}^\star = 
\underset{\vect{\alpha}}{\arg \min} 
\| \alpha \|_{1} \:\: \mbox{such that} \:\:
\| \vect{y} - \sensmat \sparmat \vect{\alpha}\|_2 \leq \epsilon
\spcend ,
\end{equation}
where the image is synthesising by $\imvect^{\star} = \sparmat
\vect{\alpha}^\star$, or by solving the Total Variation (TV) problem
\begin{equation}
\label{eqn:min_tv}
\imvect^{\star} = 
\underset{\vect{x}}{\arg \min}
\| \vect{x} \|_{\rm TV} \:\: \mbox{such that} \:\:
\| \vect{y} - \sensmat \vect{x}\|_2 \leq \epsilon
\spcend ,
\end{equation}
respectively.  Recall that the $\el_1$-norm $\| \cdot \|_{1}$ is
simply given by the sum of the absolute values of the elements of a
vector and the squared $\el_2$-norm $\| \cdot \|_{2}^2$ is given by
the sum of the squares of the elements of a vector.  The TV norm $\|
\cdot \|_{\rm TV}$ is given by the $\el_1$-norm of the gradient of the
signal.  The tolerance $\epsilon$ is related to an estimate of
the noise variance.


To extend the standard compressed sensing imaging framework to wide
fields \cite{mcewen:2010:riwfov}, interferometric images are
considered directly on the sphere, rather than the equatorial plane.
The measurement operator \sensmats\ transforming the image defined on
the celestial sphere $\imsvect$ to measurements $\visvect$, consists
of augmenting the usual interferometric measurement operator with an
initial projection $\opproj$ from the sphere to the plane, \ie\
\begin{equation}
\label{eqn:vis_linear_sphere}
\visvect = \sensmats \imsvect + \nimvect
\spcend ,
\end{equation}
where $\sensmats = \sensmatp \opproj$.
The initial projection simply corresponds to a change from spherical
to Cartesian coordinates, resulting in a framework which remains
general and does not rely on any small-field assumptions.  However, the
projection which implements the change of variable is complicated by
the discrete setting and the desire to recover a regular grid on the
plane to allow the use of fast Fourier transforms (FFTs).  In order to
project onto a regular grid on the plane, it is necessary to re-grid
the pixelisation on the sphere to recover sample values at spherical
positions that project directly onto the planar grid.
A convolution on the sphere is
incorporated in the projection operator to achieve this re-gridding.
The convolutional re-gridding on the sphere is similar to the
re-gridding often performed when mapping the measurements observed by
an interferometer at continuous coordinates to a regular grid, also
to afford the use of FFTs \citep{thompson:2001}.  Careful
consideration is also given to samplings on the sphere and plane to
ensure that the planar grid is sampled sufficiently to accurately
represent the projection of a band-limited signal defined on the
sphere.  Spherical interferometric images may then be recovered by
solving the optimisation problems given by \eqn{\ref{eqn:min_bp}} and
\eqn{\ref{eqn:min_tv}}, by replacing the measurement operator
\sensmatp\ with its spherical equivalent \sensmats.

The performance of compressed sensing reconstruction is driven by two
factors: sparsity and coherence.  Both of these factors can be
enhanced in the wide-field spherical interferometric imaging framework
\cite{mcewen:2010:riwfov}.  
A signal $\vect{x}$ is said to be sparse if there exists a sparsifying
basis yielding coefficients $\vect{\alpha} = \sparmat^{\rm T} \imvect
$, for which the number of non-zero coefficients is much smaller than
the dimensionality of the original signal.  The theory of compressed
sensing states that the more sparse a signal the fewer measurements
required to recover it, or similarly, the better the reconstruction
quality for a given number of measurements.  By recovering
interferometric images on the sphere, distorting projections are
eliminated and the sparsity of the signal is enhanced, improving the
performance of compressed sensing reconstruction.
Coherence between the measurement and sparsity bases is also a
critical factor driving reconstruction performance: as the coherence
between the two bases increases, the reconstruction performance
degrades.  Incoherence ensures that the measurement basis $\sensmat$
cannot sparsely represent the sparsity basis $\sparmat$, ensuring that
signal content is sufficiently probed by incomplete measurements.
The so-called spread spectrum phenomenon \cite{wiaux:2009:ss} arises
by relaxing the small-field assumption of standard interferometric
imaging, resulting in a non-negligible \bw\ component modulation in
the measurement operator.  This modulation may be seen as a
convolution of the Fourier representation of the image, spreading its
spectrum and increasing incoherence with the (essentially) Fourier
measurement basis.  The greater the frequency content of the
modulation, the larger the spreading.  Since the maximum frequency
content of the modulation increases with the field-of-view, the spread
spectrum phenomenon is more effective at improving reconstruction
quality the wider the field-of-view.
Note that the universality and efficiency of the spread spectrum
technique was recently demonstrated on purely theoretical grounds
beyond its application to radio interferometry \cite{puy:2011}.
The wide-field interferometric imaging framework thus provides
intrinsic advantages, enhancing both sparsity and incoherence, and,
consequently, the fidelity of reconstructed images.

\newlength{\fdsplotwidth}
\setlength{\fdsplotwidth}{51mm}

\begin{figure}[t]
\centering
\mbox{
\subfigure[Ground truth]{\quad \includegraphics[clip=,viewport=45 30 367 320,width=\fdsplotwidth]{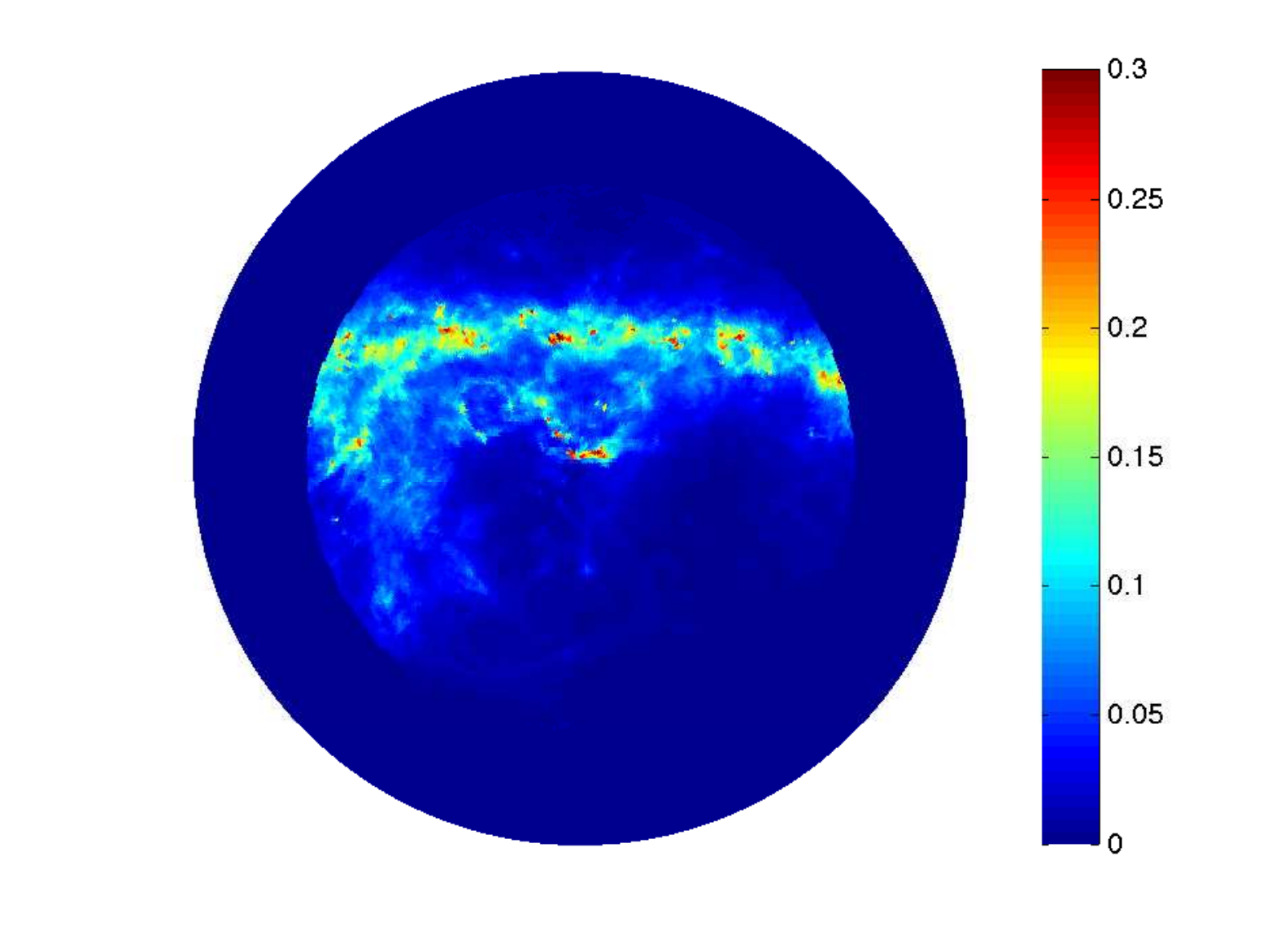} \quad} 
\subfigure[Planar reconstruction with SS]{\quad \includegraphics[clip=,viewport=45 30 367 320,width=\fdsplotwidth]{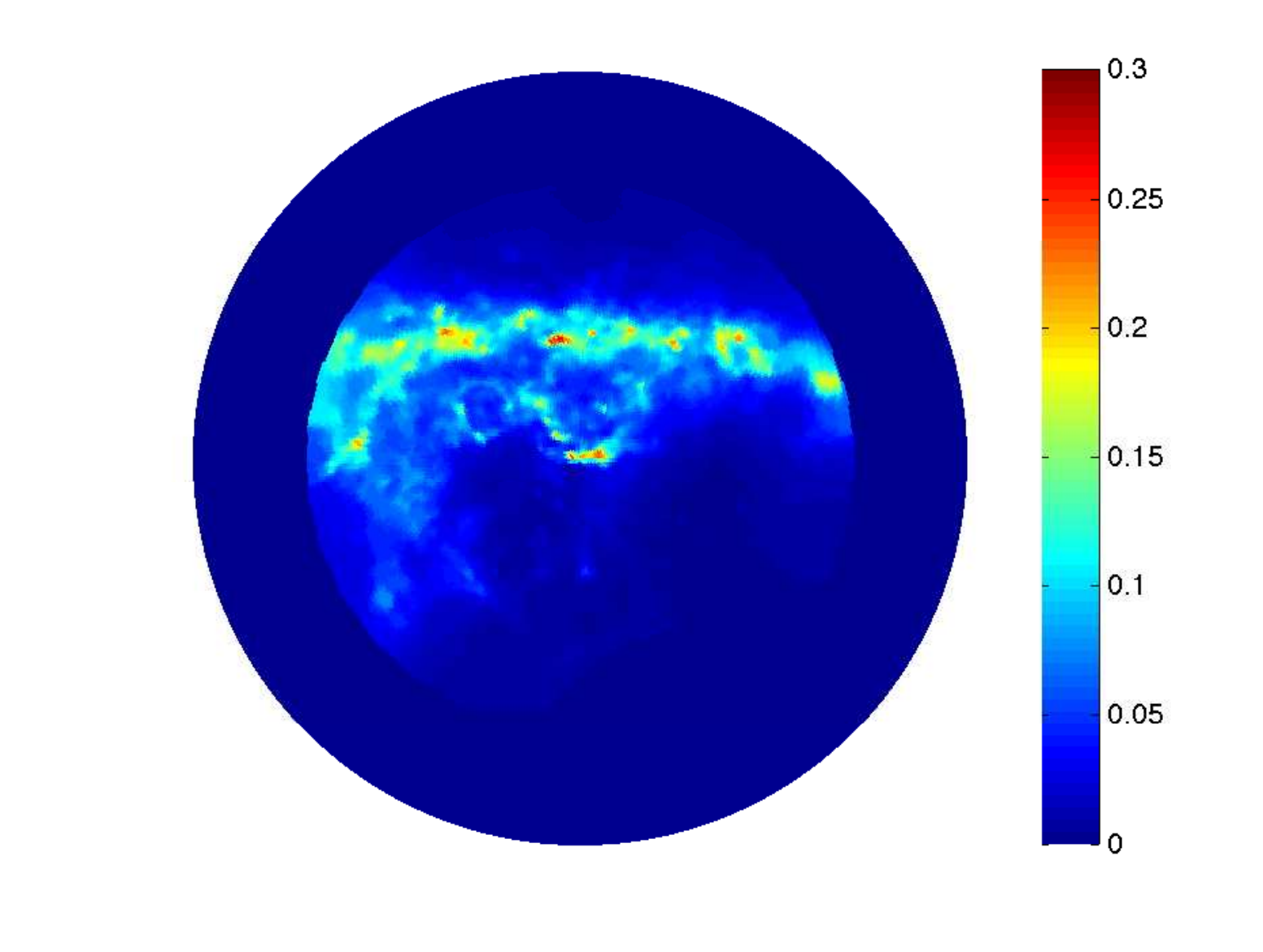} \quad}
}\\
\mbox{
\subfigure[Spherical reconstruction without SS]{\quad \includegraphics[clip=,viewport=45 30 367 320,width=\fdsplotwidth]{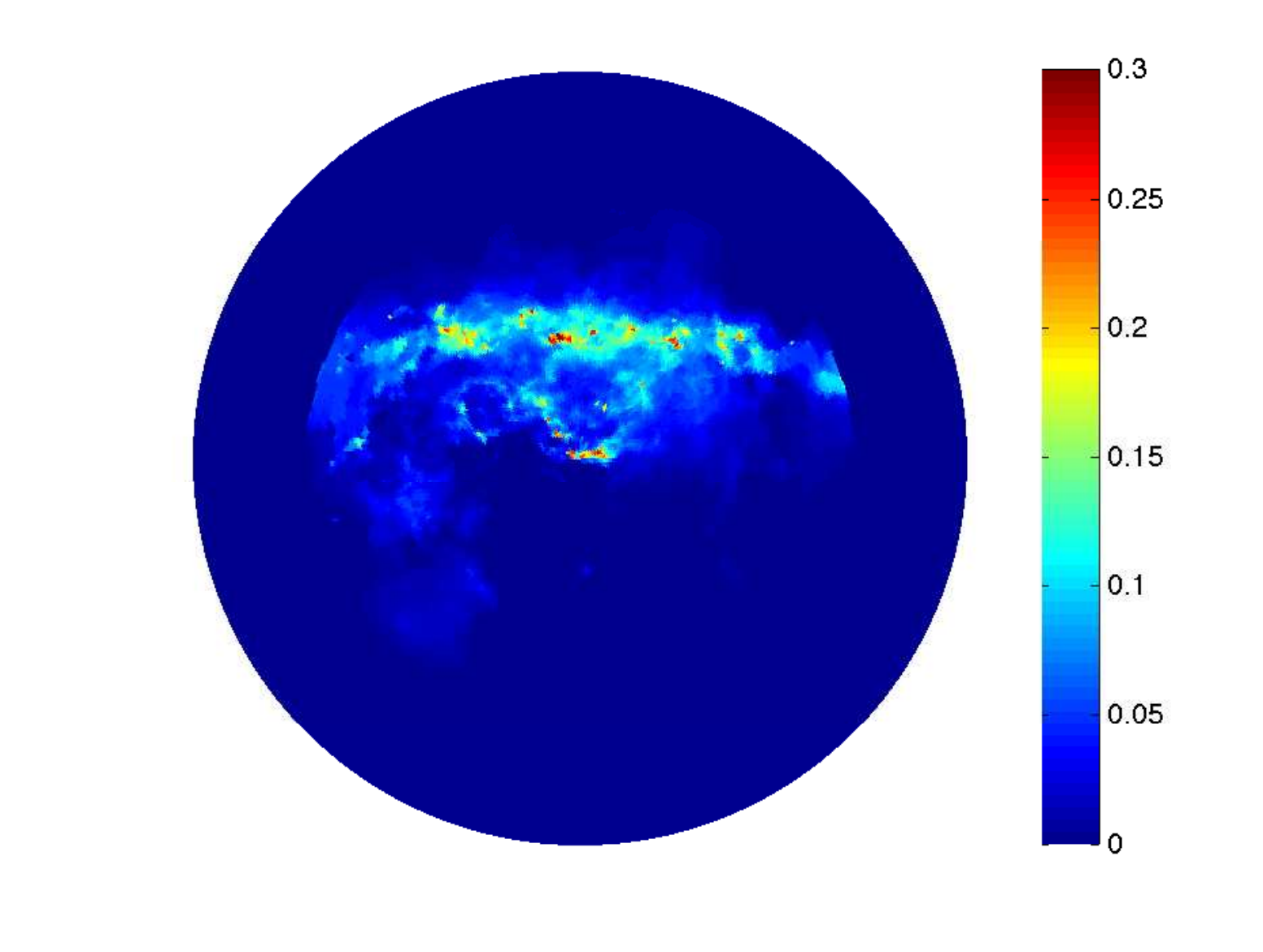} \quad}
\subfigure[Spherical reconstruction with SS]{\quad \includegraphics[clip=,viewport=45 30 367 320,width=\fdsplotwidth]{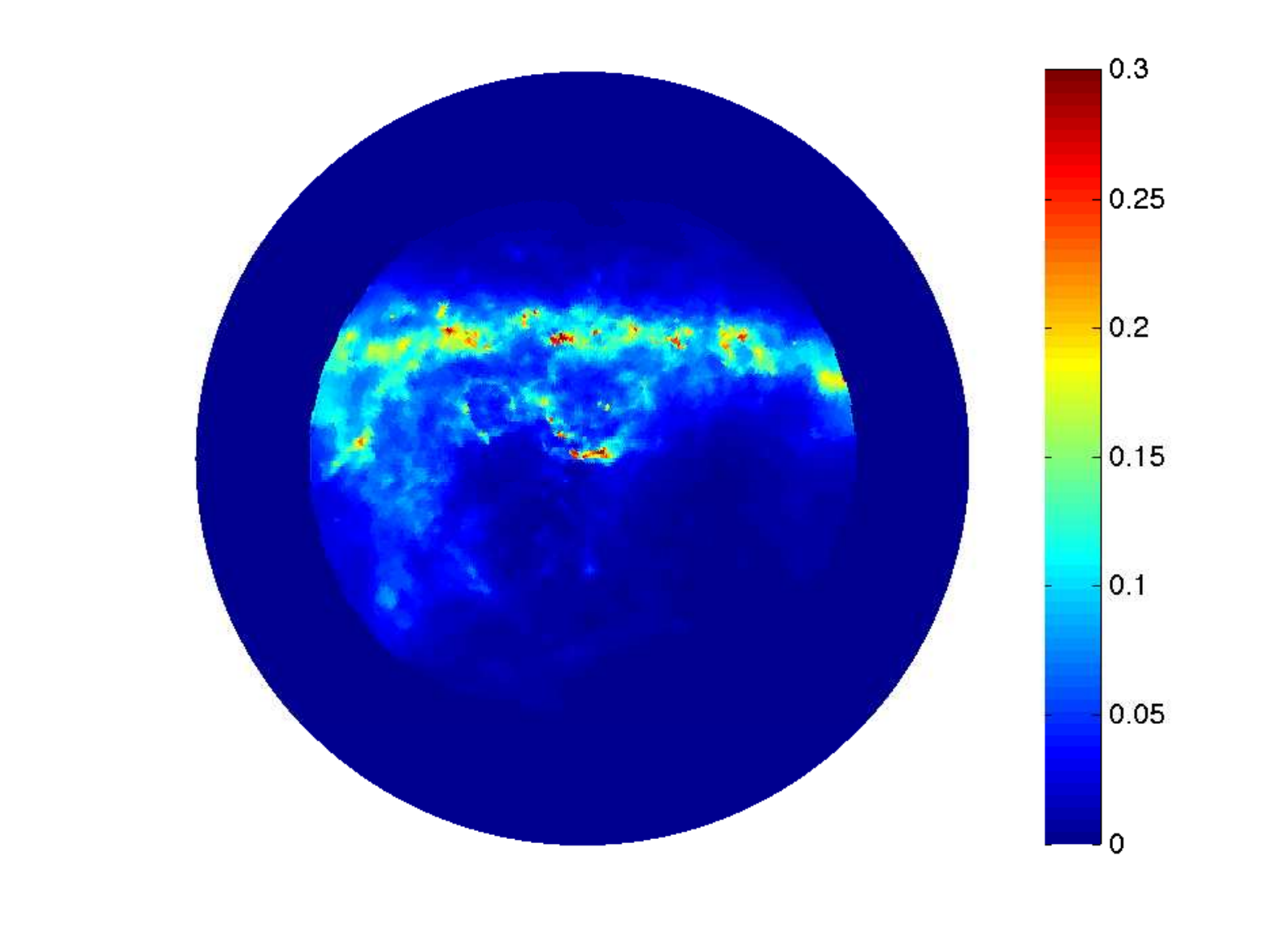} \quad}
}
\caption{Simulated reconstructions of diffuse interstellar Galactic
  dust in the presence and absence of the spread spectrum (SS) phenomenon.}
\label{fig:fds}
\end{figure}

\section{Simulations}

The wide-field interferometric imaging framework has been evaluated
thoroughly on low-resolution simulated observations of sources with a
Gaussian profile, where a direct comparison with planar
reconstructions was made \cite{mcewen:2010:riwfov}.  The predicted
improvement in the fidelity of reconstructed images in the wide-field
setting, due to the theoretical considerations discussed previously, is
indeed realised in practice.
A more realistic simulation of Galactic dust emission
\citep{finkbeiner:1999} at a higher resolution was also considered
\cite{mcewen:2010:riwfov}.  In \fig{\ref{fig:fds}} we plot the
underlying spherical image of this simulation, and images
reconstructed in the planar and spherical compressed sensing
frameworks, with only 25\% of Fourier samples measured (for
comparison, the planar reconstruction is lifted to the sphere -- the
space where the image naturally lives).  These images are recovered by
solving the TV problem.  The signal-to-noise-ratio (SNR) of the
spherical reconstruction in the absence of the spread spectrum
phenomenon is $7$dB, while the reconstructed images in the presence of
the spread spectrum phenomenon is $14$dB and $19$dB for the planar and
spherical reconstructions respectively.

\section{Conclusions and future}

The intrinsic advantages of the \bw\ component and spherical imaging
for wide-field radio interferometry in the context of compressed
sensing are clear.  However, current techniques are necessarily
somewhat idealise in order to remain as close as possible to the
theoretical compressed sensing setting. Now that the effectiveness of
these techniques has been demonstrated, it is of paramount importance
to adapt them to realistic interferometric configurations.
Furthermore, the possibility of optimising the configuration
of interferometers to enhance the spread spectrum phenomenon for
compressed sensing reconstruction is an exciting avenue of research at
the level of acquisition.
In summary, next-generation radio interferometric telescopes, such as
the SKA, will inherently observe very large fields of view.  Enhanced
wide-field interferometric imaging techniques are therefore of increasing
importance to ensure that the fidelity of reconstructed images keeps
pace with the capabilities of new instruments.

\section{Acknowledgments}

JDM is supported by the Swiss National Science Foundation (SNSF) under
grant 200021-130359.  YW is supported by the Center for
Biomedical Imaging (CIBM) of the Geneva and Lausanne Universities,
EPFL, and the Leenaards and Louis-Jeantet foundations, and by
the SNSF under grant PP00P2-123438.

\setlength{\bibsep}{0.5mm}
\renewcommand{\bibsection}{\section{References}}
\bibliographystyle{IEEEbib}
\bibliography{bib}

\end{document}